\documentclass[reprint,superscriptaddress,prl,a4paper]{revtex4-2}
\usepackage[utf8]{inputenc}
\pdfoutput=1
\usepackage{mathptmx,graphicx}

\usepackage[a4paper,width=150mm,top=25mm,bottom=25mm,bindingoffset=0mm]{geometry}

\usepackage[version=3]{mhchem} 
\usepackage{soul}
\usepackage{graphicx}
\usepackage{wrapfig}
\usepackage{amsmath}
\usepackage[export]{adjustbox}
\usepackage{float}

\usepackage{textgreek}
\newsavebox{\measurebox}

\usepackage{xcolor}
\usepackage[colorlinks=true,citecolor=black,linkcolor=black]{hyperref}

\usepackage{blindtext}

\usepackage{comment} 

\usepackage{wasysym} 
\usepackage{mathtools} 
\def \tpsi {\widetilde{\psi}}
\def \tomega {\widetilde{\omega}}

\begin{document}

\title{Controlling lasing around Exceptional Points in Coupled  Nanolasers}

\author{Anna Fischer}
\affiliation{Department of Physics, Imperial College London, London SW7 2AZ, United Kingdom}
\affiliation{IBM Research Europe - Z\"{u}rich, S\"{a}umerstrasse 4, R\"{u}schlikon, 8803, Switzerland}
\author{T. V. Raziman}
\affiliation{Department of Physics, Imperial College London, London SW7 2AZ, United Kingdom}
\affiliation{Department of Mathematics, Imperial College London, London SW7 2AZ, United Kingdom}
\author{Wai Kit Ng}
\affiliation{Department of Physics, Imperial College London, London SW7 2AZ, United Kingdom}
\author{Jente Clarysse}
\affiliation{Department of Physics, Imperial College London, London SW7 2AZ, United Kingdom}
\author{Dhruv Saxena}
\affiliation{Department of Physics, Imperial College London, London SW7 2AZ, United Kingdom}
\author{Jakub Dranczewski}
\affiliation{IBM Research Europe - Z\"{u}rich, S\"{a}umerstrasse 4, R\"{u}schlikon, 8803, Switzerland}
\affiliation{Department of Physics, Imperial College London, London SW7 2AZ, United Kingdom}
\author{Stefano Vezzoli}
\affiliation{Department of Physics, Imperial College London, London SW7 2AZ, United Kingdom}
\author{Heinz Schmid}
\affiliation{IBM Research Europe - Z\"{u}rich, S\"{a}umerstrasse 4, R\"{u}schlikon, 8803, Switzerland}
\author{Kirsten Moselund}
\affiliation{Paul Scherrer Institut, Forschungsstrasse 111, Villigen, 5232, Switzerland}
\affiliation{EPFL, Lausanne, 1015, Switzerland}
\author{Riccardo Sapienza}
\affiliation{Department of Physics, Imperial College London, London SW7 2AZ, United Kingdom}

\begin{abstract}
Coupled nanolasers are of growing interest for on-chip optical computation and data transmission, which requires an understanding of how lasers interact to form complex systems. The non-Hermitian interaction between two coupled resonators, when excited selectively, can lead to parity-time symmetry, the formation of exceptional points, and subsequently spectral control and increased sensitivity. These investigations have been limited to pump energies close to the lasing threshold, and  large or narrow-line lasers. 
Here, by programmable optical excitation we study two coupled nanolasers significantly above threshold, where mode instability plays an important role. We map the mode evolution around two exceptional points, and observe lasing gaps due to reversed pump dependence which compare well with nonlinear theory. Finally, the coupling can be exploited to control the lasing threshold and wavelength, and for frequency switching around the lasing gap. 
Controlled and integrated nanolasers constitutes a promising platform for future highly sensitive and programmable on-chip laser sources. 
\end{abstract}

\maketitle

\subsection*{Introduction}

Tunable coherent nanoscale light sources are important for many technologies, ranging from on-chip optical computing \cite{brunnerCompetitivePhotonicNeural2021, shenDeepLearningCoherent2017, tirabassiBinaryImageClassification2022a} and data transmission \cite{wangNovelLightSource2017a} to sensing \cite{ zhongSensingExceptionalSurfaces2019} and biophysics \cite{caixeiroLocalSensingAbsolute2023}. 
Nanolasers are easier for integration on chip and reach GHz modulation \cite{xuDirectModulation2022}.
However, fabrication of nanolasers with a specific spectral response is challenging due to intrinsic fabrication imperfections \cite{gilsantosScalableHighprecisionTuning2017, mautheInPonSiOpticallyPumped2019a, tiwariSingleModeEmissionInP2022, dranczewskiPlasmaEtchingFabrication2023}.
The challenges are exacerbated when many nanolasers are made to interact, to 
increase power output and functionality, due to the complex collective effects \cite{choiRoomTemperatureElectrically2021, qiaoHigherdimensionalSupersymmetricMicrolaser2021, lepriComplexActiveOptical2017a, tirabassiBinaryImageClassification2022a}.

In coupled nanolasers, parity-time symmetry and its spontaneous breaking at the exceptional point (EP), where eigenmodes of the system coalesce, 
\cite{benderRealSpectraNonHermitian1998,  ozdemirParityTimeSymmetry2019, el-ganainyNonHermitianPhysicsPT2018, hamelSpontaneousMirrorsymmetryBreaking2015, miriExceptionalPointsOptics2019} can be exploited as a powerful tool for post-fabrication spectral control. 
Selectively pumped coupled microdisk lasers are interesting as they provide a wide range of functionalities such as single mode emission \cite{fengSinglemodeLaserParitytime2014, hodaeiParitytimeSymmetricMicroring2014, hodaeiParitytimesymmetricCoupledMicroring2015, wangDualwavelengthSinglefrequencyLaser2016, hodaeiDesignConsiderationsSingleMode2016, liuSingleModeLasingSpontaneous2023}, 
optical isolation \cite{pengParityTimesymmetricWhisperinggallery2014, changParityTimeSymmetry2014}, 
chiral emission \cite{hayengaDirectGenerationTunable2019}, 
and sensing \cite{chenExceptionalPointsEnhance2017, wiersigEnhancingSensitivityFrequency2014, zhongPowerlawScalingExtreme2018, wiersigReviewExceptionalPointbased2020}.
Moreover, the dependence of lasing on gain can be reversed near EPs, resulting in suppression and  revival of lasing  \cite{liertzerPumpInducedExceptionalPoints2012, brandstetterReversingPumpDependence2014, pengLossinducedSuppressionRevival2014}.

Yet, the lasing characteristics around EPs have so far mostly been studied
for pump energies close to the lasing threshold, where mode competition
is minimal and linear coupled mode theory applies~\cite{hodaeiParitytimeSymmetricMicroring2014, hodaeiDesignConsiderationsSingleMode2016}. 
Instead, operation well above threshold is sought for deeper lasing understanding \cite{benzaouiaNonlinearExceptionalpointLasing2022a} and to exploit new phenomena such as chaotic behaviours and instabilities \cite{jiTrackingExceptionalPoints2022}.

 Experiments with large microdisks ($10-100$~$\mu$m radii) \cite{fengSinglemodeLaserParitytime2014, hodaeiParitytimeSymmetricMicroring2014, hodaeiParitytimesymmetricCoupledMicroring2015, wangDualwavelengthSinglefrequencyLaser2016, hodaeiDesignConsiderationsSingleMode2016, liuSingleModeLasingSpontaneous2023, pengParityTimesymmetricWhisperinggallery2014, changParityTimeSymmetry2014,
 hayengaDirectGenerationTunable2019,
 chenExceptionalPointsEnhance2017}, or photonic crystal cavities \cite{hamelSpontaneousMirrorSymmetry2015, jiTrackingExceptionalPoints2022} achieve coupling more easily due to narrow linewidths. Instead, in devices of sizes comparable to the wavelength, 
the effect of coupling is reduced by the low Q-factors, limiting the study of lasing around EPs~\cite{ishiiBistableLasingTwin2005, nakagawaPhotonicMoleculeLaser2005}. 

Here, we study nanoscale coupled InP microdisks, bottom-up grown by epitaxy \cite{staudingerWurtziteInPMicrodisks2020a}, when excited well above (2x) threshold by selective and programmable illumination.
We map the two virtual EPs of the system, and observe opening of lasing gaps due to reversed pump dependence at the EPs. We demonstrate emission tunability and switching. 
The experimental findings are confirmed by non-linear coupled mode theory that includes gain-saturation and stability analysis.
\\

\subsection*{Results}

\begin{figure}[ht!]
    \centering
    \includegraphics[width=\linewidth]{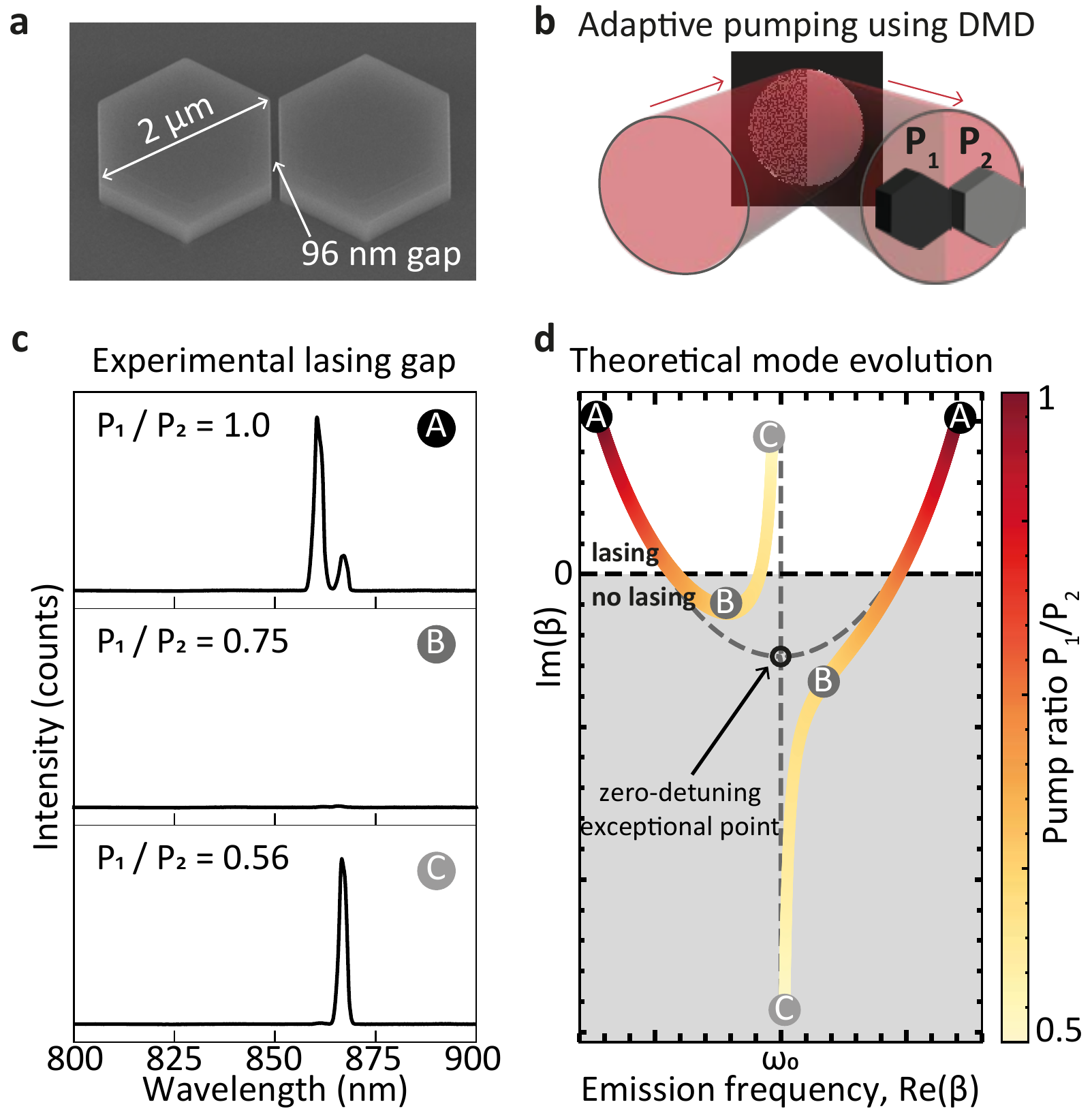}
    \caption{\textbf{Counter-intuitive lasing behaviour in coupled microdiks lasers}
    (a) SEM image of two coupled bottom-up grown InP microdisk lasers with 2 \textmu{}m diameter and 96 nm gap. 
    (b) Schematic of the DMD setup and pattern used for selective pumping, where the uniform beam is reflected from the DMD and shaped into two half circles with powers $P_1$ and $P_2$.
    (c) Three emission spectra illustrating the lasing gap under selective pumping of coupled InP microdisk lasers. $P_2$ = 163 \textmu{}J cm$^{-2}$ pulse$^{-1}$ for all three acquired spectra while $P_1$ is increased from bottom to top. 
    (d) Evolution of real and imaginary parts of mode eigenvalues based on linear coupled mode theory under selective pumping. In the experimental spectrum, the modes have different intensities due to effects such as mode competition and non-equal losses of the disks, which are not included in coupled mode theory.}
    \label{fig:1}
\end{figure}

The coupled InP microdisk lasers are bottom-up grown, ensuring a defect-free crystal structure with extremely smooth side walls corresponding to the $\left\langle111\right\rangle$ facets and high-quality gaps \cite{staudingerWurtziteInPMicrodisks2020a, schmidTemplateassistedSelectiveEpitaxy2015}. This unique platform reduces cavity losses and makes it possible to reach efficient coupling.
The microdisk lasers support few resonant modes within the gain spectral range, owing to their small size, with diameters ranging between 0.5 -- 2 \textmu{}m and heights between 0.4 -- 1 \textmu{}m, respectively.

The InP microdisks are characterised in a micro photoluminescence and lasing setup. They are excited with a 633~nm, 200~fs-pulsed pump laser. Individual microdisks show single-mode emission with emission wavelengths between 800 and 900 nm, strongly depending on their diameter and height, and with thresholds in the range of 100 - 700 \textmu{}J~cm$^{-2}$~pulse$^{-1}$ (Figure S2).
We estimate the Q-factor to be below 300 similar to previous works~\cite{mautheInPonSiOpticallyPumped2019a}. 
The low Q-factor originates from the small size and leads to a stronger evanescent field outside the disks, which enhances the coupling and enables measurable effects for inter-laser distances up to $\sim \lambda / 6$.
The well-defined gap with smooth side-walls ensures optimal coupling (Figure~\ref{fig:1}a).

We excite the InP microdisk lasers with variable pump on the two disks, that we control using a DMD \cite{saxenaSensitivitySpectralControl2022}. The beam is shaped into two half circles with programmable pumping powers $P_1$ and $P_2$ (Figure~\ref{fig:1}b). The pump power per half circle is varied by defining the fraction of reflecting pixels using a dither function (see Methods). 
This is an improvement over previous methods such as covering parts of the pump beam with a knife edge \cite{hodaeiParitytimesymmetricCoupledMicroring2015}, notch filter \cite{hodaeiParitytimeSymmetricMicroring2014} or simply moving the beam off the device \cite{liuSingleModeLasingSpontaneous2023}, which are inhomogeneous and difficult to control.

The system exhibits a counter-intuitive reversal of the pump dependence of lasing when we keep the pump on disk 2 ($P_2$) constant and increase the pump $P_1$ on the other (Figure~\ref{fig:1}c). 
The spectrum evolves, with increasing pump ratio, from single mode lasing ($P_1/P_2$ = 0.56), to no lasing ($P_1/P_2$ = 0.75), to multimode lasing ($P_1/P_2$ = 1). 
This observation of reversed pump dependence proves that the microdisks are coupled efficiently.
The symmetry of the system ensures the same behaviour when $P_1$ is kept constant and $P_2$ varies.

The reversed pump dependence can be well captured by coupled mode theory (see Methods) as shown in Figure~\ref{fig:1}d, where each point A-C corresponds to an experimental plot in Figure~\ref{fig:1}c. 
If both microdisks are excited equally (A), the modes of the system split in their real parts (the frequency) while keeping the same imaginary part (the gain/loss), resulting in both modes lasing. When $P_1$ is significantly less than $P_2$ (C), the modes are predominantly split in imaginary part, resulting in single-mode emission. For intermediate values of $P_1/P_2$, the two modes approach each other in imaginary part
, with both  imaginary parts going below threshold for a range of ratios (B). For a zero-detuned system (gray dashed line), where $\omega_1 = \omega_2 = \omega_0$, the two modes meet at an exceptional point (EP), where both the eigenvalues and eigenvectors coalesce (identified by the arrow in Figure~\ref{fig:1}d)~\cite{benderRealSpectraNonHermitian1998,  ozdemirParityTimeSymmetry2019, el-ganainyNonHermitianPhysicsPT2018, hamelSpontaneousMirrorsymmetryBreaking2015, miriExceptionalPointsOptics2019}.
At the EP, the system is highly sensitive to perturbations showing strong responses in the spectral features~\cite{chenExceptionalPointsEnhance2017, wiersigEnhancingSensitivityFrequency2014, zhongPowerlawScalingExtreme2018, wiersigReviewExceptionalPointbased2020}.
Slight detuning of the resonant frequencies (coloured line) $\omega_i = \omega_0 \pm \delta$ produces an avoided crossing around the zero-detuning EP, which still has similar implications on the spectral response as an EP.

\begin{figure*}[ht!]
    \centering
    \includegraphics[width= \linewidth]{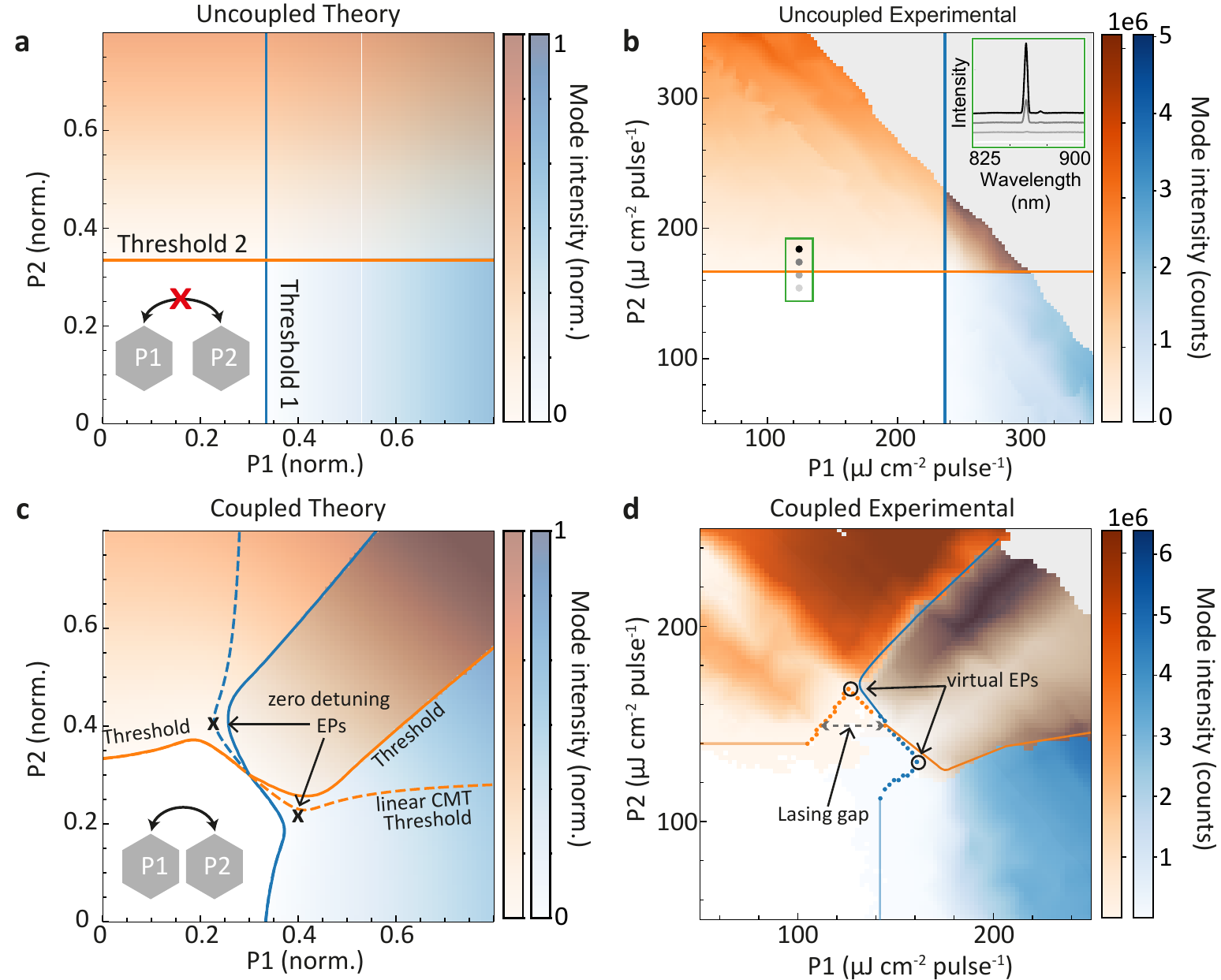}
    \caption{\textbf{Mode thresholds become highly co-dependent in a coupled system} 
    (a) Theoretical mode landscape as a function of pumping power $P_{1,2}$ of two \textit{un}coupled microdisks (schematic in inset) showing how the threshold of each disk is independent of pumping on the other disk and only dependent on the intrinsic losses which are set to $\gamma_1 = \gamma_2 = \gamma = 0.3$.
    (b) Experimental mode landscape of two uncoupled InP microdisks with 2 \textmu{}m diameter and 680 nm gap showing the independent thresholds of the two microdisks. The green inset shows emission spectra at four pump powers (fixed P$_1$ = 74 \textmu{}J cm$^{-2}$ pulse$^{-1}$; P$_2$ = 154, 164, 174, and 184 \textmu{}J cm$^{-2}$ pulse$^{-1}$), indicated as grayscale dots in main plot.
    (c) Theoretical mode landscape of two coupled microdisks (schematic in inset) with slight frequency detuning $\delta = 0.02$, coupling $\kappa = 0.1$, and losses $\gamma = 0.3$. The threshold lines of the two modes are clearly affected by increased pump power on the other disk. The two \textbf{x} mark where EPs would lie in a zero-detuned system. Due to detuning, avoided crossings are visible around the EPs. An instability region separates the theoretical threshold (dashed line) from the actual onset of the second mode (solid lines).
    (d) Experimental mode landscape of two coupled InP microdisks with 2 \textmu{}m diameter and 96 nm gap. The indicated threshold lines are a combination of calculated values (dots) and estimated visual guides (lines). The coupling results in two virtual EPs around which lasing gaps open (gray dashed arrow). 
   \textbf{For all plots}, the intensity of the mode is given by the colour (see colour bars). White: below threshold, grey area: no data.}
    \label{fig:2}
\end{figure*}

We can obtain a complete map of the lasing landscape including the behaviour far above threshold by varying $P_1$ and $P_2$  independently (Figure~\ref{fig:2}). 
The intensities of mode 1 (blue shading) and mode 2 (orange shading) are shown for both uncoupled (top panels) and coupled (bottom panels) systems, based on theoretical predictions (left panels) and experimental results (right panels).

When two resonators are \emph{uncoupled}, each laser is unaffected by the pump on the other disk (Figure~\ref{fig:2}a). Therefore, the system has one lasing mode per disk.
A mode reaches threshold when it is pumped sufficiently to offset the losses, making the threshold lines (blue and orange lines) parallel to the pump axes.
Lasing measurements of uncoupled InP microdisks reproduce the theoretical threshold behaviour (Figure~\ref{fig:2}b, for 2 \textmu{}m diameter, 680 nm gap device). 
The lasing spectrum is shown in the inset of Figure~\ref{fig:2}b for different pump powers (dots in the main figure).

When two resonators are \emph{coupled}, the modes become delocalised across the two disks, which makes lasing action dependent on both pump powers $P_{1,2}$ and the threshold lines are no longer straight lines aligned along the pump axis (Figure~\ref{fig:2}c,d). 
We first analyse the lasing action for pump powers close to the lasing threshold with linear coupled mode theory (see Methods).
We calculate the theoretical modes for resonators with slight frequency detuning ($\delta / \kappa = 0.2$) to better reflect the experimental conditions (see later). 
When the pump powers on both disks become 
close to the lasing threshold, the coupling transfers gain from the more highly pumped disk to the other.
As a result, the threshold of the more pumped disk increases with respect to the uncoupled value while that of the less pumped disk decreases, pulling the threshold lines towards each other (solid lines in Figure~\ref{fig:2}c).
In a zero-detuned system, the threshold lines would merge into a single segment, known as the parity-time symmetry line \cite{ozdemirParityTimeSymmetry2019}, where both modes have the same threshold (Figure S5).
At the ends of the parity-time symmetry line lie the EPs of the zero-detuned system, where the eigenmodes coalesce. Frequency detuning prevents this coalescence and instead creates avoided crossings around the zero-detuning EPs (black crosses in Figure~\ref{fig:2}c).

\begin{figure*}[t!]
    \centering
    \includegraphics[width = \linewidth]{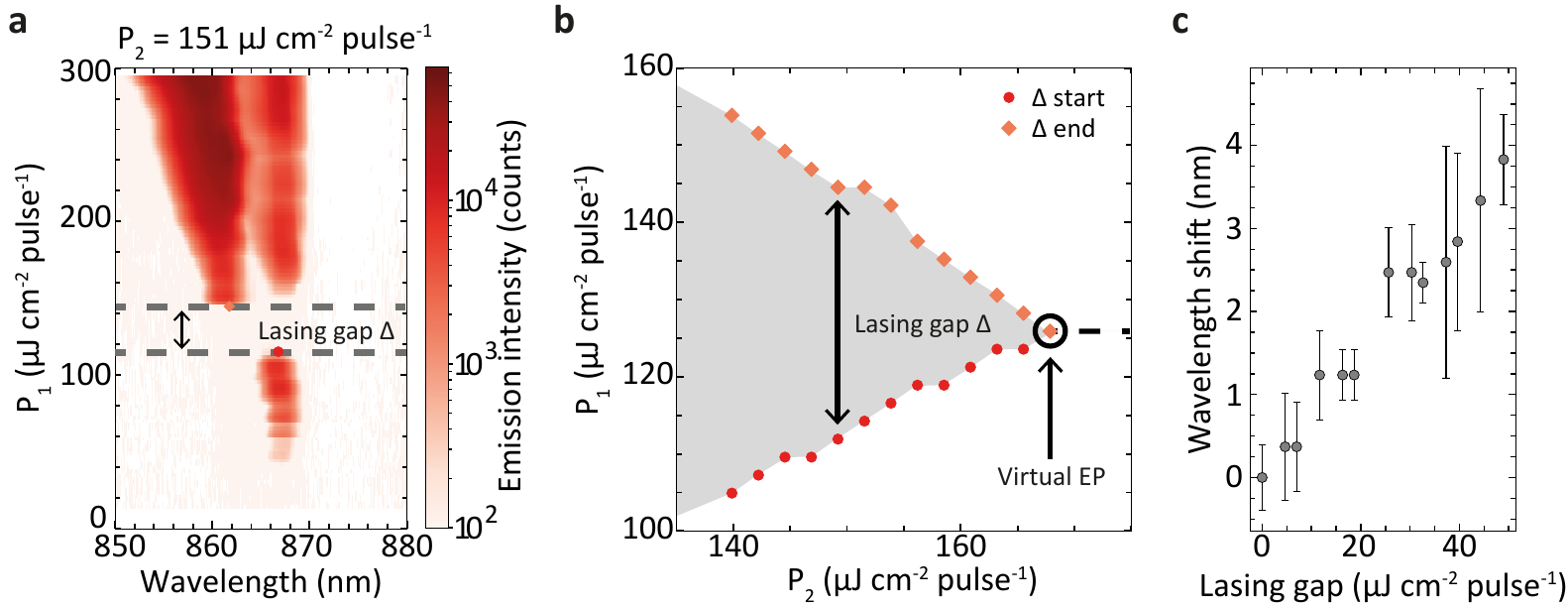}
    \caption{\textbf{The lasing gap narrows under increased P$_1$ and closes at a "virtual" EP.}
    (a) Pump power on disk 1 ($P_1$) versus emission wavelength for constant $P_2$ of 151 \textmu{}J cm$^{-2}$ pulse$^{-1}$. The colourbar indicates the emission intensity. While $P_1$ is increased, the lasing stops and then resurfaces at a shifted frequency resulting in a counter-intuitive lasing gap.
    (b) The lasing gap decreases under increased $P_2$. When the lasing gap closes, a "virtual" (due to experimentally inherent detuning) EP at the avoided crossing is reached. 
    (c) Wavelength shift across the lasing gap versus lasing gap size.}
    \label{fig:3}
\end{figure*}

Above threshold, gain saturation and mode competition modify the lasing behaviour.
This can be captured by extending linear coupled mode theory to include gain-saturation \cite{benzaouiaNonlinearExceptionalpointLasing2022a, santosGainClampingRandom2021} and stability analysis \cite{strogatzNonlinearDynamicsChaos2019}. 
With this non-linear model, we compute the mode intensities above threshold. (see Supplementary Information section SIII).
When one mode is above threshold, the second mode (solid line, Figure~\ref{fig:2}c) appears at a higher threshold than predicted by linear coupled mode theory (dashed line).
Between the solid and dashed lines, the second mode is unstable and therefore only the mode with the higher gain lases. 

Experimental threshold lines of coupled InP microdisks follow the features predicted by theory (Figure~\ref{fig:2}d, for 2 \textmu{}m diameter, 96 nm gap device).
The indicated threshold lines are a combination of calculated values (orange and blue dots) and guides (orange and blue lines). 
When the pump power is similar in both disks, the effect of increase in threshold, with a similar shape as predicted by theory, is observed (orange and blue dots in Figure~\ref{fig:2}d).
The threshold lines lie nearly on top of each other in this region due to the large coupling between the disks.
The black circles indicate the position of  so-called ``virtual" EPs. These points are the minima of the avoided crossings at the zero-detuning EPs and represent the closest a physical system can come to an EP without active detuning modulation \cite{jiTrackingExceptionalPoints2022}.
The experimental onset of the second lasing mode above threshold follows the theory and confirms the need for non-linear coupled mode theory above threshold. The threshold of mode 1 (blue line) for a device already lasing in mode 2 (orange shading) follows a straight line with positive slope above threshold. At very high pumping powers, $P_2 > 250$ \textmu{}J cm$^{-2}$ pulse$^{-1}$, the threshold of mode 2 (orange line) deviates from the theoretical prediction, which we attribute to non-linear interactions with higher-order modes in the system.

The virtual EPs allow us to estimate the coupling in the system, as their distance is equal to the coupling factor $\kappa = 18$~\textmu{}J cm$^{-2}$ pulse$^{-1}$ and $\kappa / \gamma = 0.13$, where $\gamma$ are the losses of the microdisks and equal to the threshold $P_{th} \approx 140$~\textmu{}J cm$^{-2}$ pulse$^{-1}$. Using the Q factor of the device, the coupling constant could be given in THz and the proportion of mode-splitting due to coupling versus detuning obtained. 
Estimates can be found in the Supporting Information (section SV).

If the coupling is larger than the laser linewidth, a lasing gap, where the pump dependence is reversed \cite{liertzerPumpInducedExceptionalPoints2012, brandstetterReversingPumpDependence2014, pengLossinducedSuppressionRevival2014}, opens near the virtual EPs (Figure~\ref{fig:2}d, dashed line).

The corresponding spectra in Figure~\ref{fig:3}a clearly show the reversal in pump dependence, where the lasing stops even though the excitation power increases. This is followed by the lasing gap and a revival of the lasing 
at shifted frequencies (Figure~\ref{fig:3}a).

The lasing gap $\Delta$, defined as the pump separation between the two lasing actions, narrows for increased $P_2$ powers and when it closes, for a pump $P_2 = 186$ \textmu{}J cm$^{-2}$ pulse$^{-1}$, the system reaches the ``virtual" EP  as shown in Figure~\ref{fig:3}b.  
The large coupling between the microdisks, which is the result of narrow and high quality gaps between the disks, enables to access large lasing gaps of up to 30\% of the lasing threshold (Figure~\ref{fig:3}b).

\begin{figure*}[t!]
    \centering
    \includegraphics[width= 0.75\linewidth]{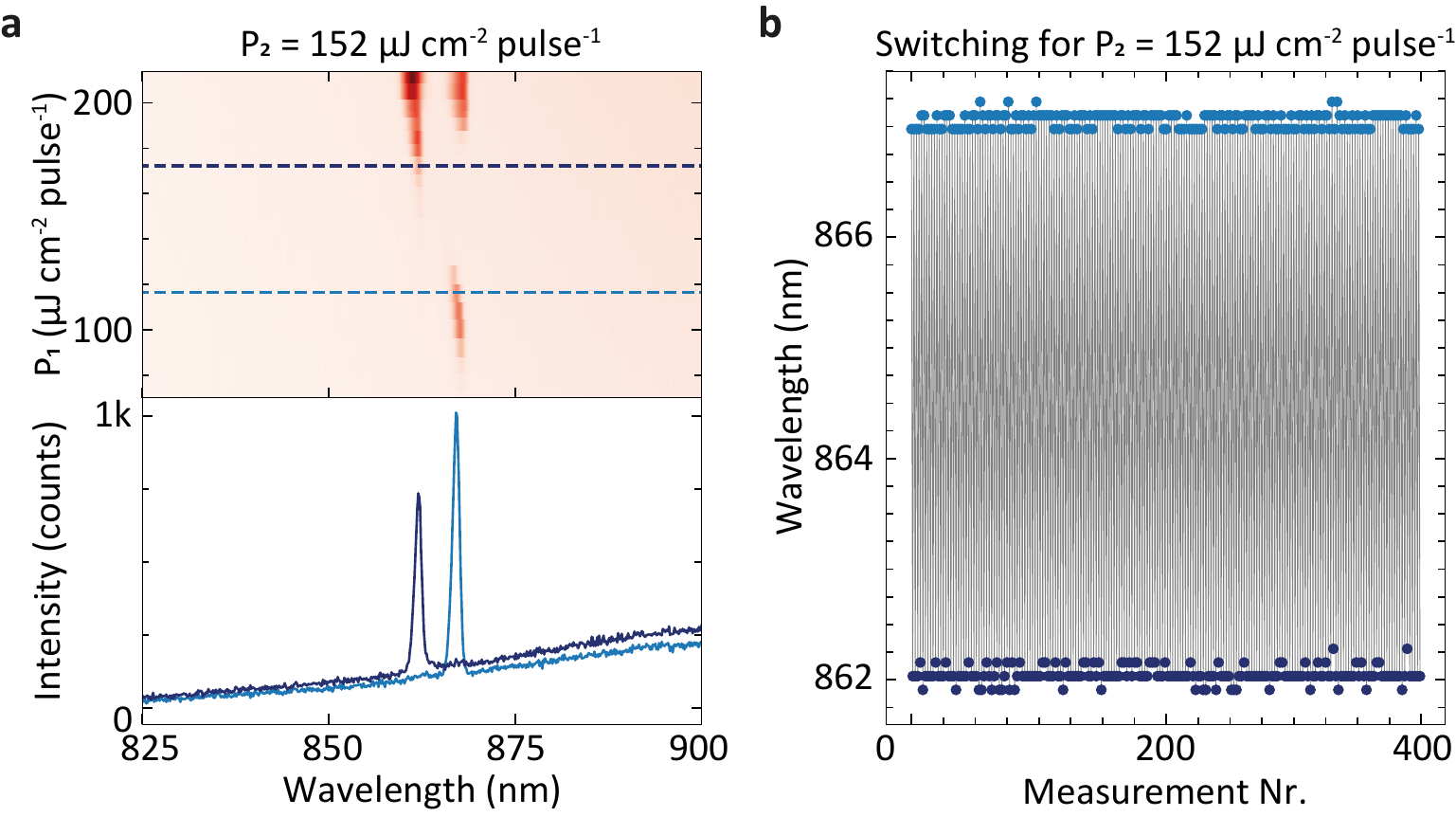}
     \caption{\textbf{The lasing gap enables stable, low power wavelength switching.}
     (a) The upper colourmap shows P$_1$ versus emission wavelength plot for constant P$_2$ of 152 \textmu{}J cm$^{-2}$ pulse$^{-1}$, with the darkness of the red indicating the emission intensity. Two spectra before (light blue, $P_1 = 117$ \textmu{}J cm$^{-2}$ pulse$^{-1}$) and after (dark blue, $P_1 = 172$ \textmu{}J cm$^{-2}$ pulse$^{-1}$) the lasing gap are shown in the lower plot and their position in the upper plot is indicated by dashed lines. 
     (b) Low power (0.8 \textmu{}W) and stable switching between two emission wavelength around the lasing gap with $\Delta \lambda$ = 5 nm and $\Delta \lambda$ / FWHM = 1.7. 
     (c) Wavelength shift versus lasing gap size obtained from the lasing gaps shown in (b).
    }
    \label{fig:4}
\end{figure*}

Around the lasing gap, the emission wavelength shifts, as the mode eigenvalues start to split in their real part. This wavelength shift is proportional to the lasing gap size and increases for larger gaps (Figure~\ref{fig:3}c). The gradient of this relation is depending on the coupling strength, of the system. For stronger coupling, the wavelength shifts more around the lasing gap as the splitting of the real parts of the eigenvalue is enhanced.

When keeping $P_2$ at 152 \textmu{}J cm$^{-2}$ pulse$^{-1}$, a wavelength shift of 3 nm is observed between spectra slightly before (light blue in Figure~\ref{fig:4}a) and after (dark blue) the lasing gap. As a reference, this shift is 10 times larger than a shift due to carrier plasma effect at similar pump powers and also ensures similar output intensities (Figure S8) \cite{bennettCarrierinducedChangeRefractive1990}.

The shift close to the virtual EP can be exploited to rapidly switch the laser emission frequency. 
We demonstrate low power (0.97(2) \textmu{}W) switching with $\Delta\lambda$ = 5~nm for a lasing peak with 1~nm FWHM, as plotted in Figure 4b. The switching is very stable with a standard deviation of only 0.07~nm, that is mostly attributed to fluctuations in the pump laser.

The pump power regime chosen for Figure~\ref{fig:4}b results in intensity modulation of 9.2 (10.5) dB for the lower (higher) wavelength peak, for the application of a small $\Delta$P around 1 \textmu{}W. This value can be significantly increased by optimising the P$_1$ and P$_2$ powers.

\subsection*{Conclusion}

To conclude, we studied bottom-up grown coupled InP microdisk lasers 
through above-threshold and selective lasing spectroscopy.
Programmable illumination allows for precise control of the lasing mode behaviour around two virtual EPs and to access functionalities such as single-mode emission, reversed pump dependence, increased sensitivity, and switching, in a single programmable device. 
The resulting lasing operation is well explained by non-linear coupled mode theory that includes gain saturation and stability analyses.
The very low-power wavelength switching around the lasing gap is promising for on-chip applications in information processing. It could also enable fast switching as it does not rely on ns-slow non-radiative recombination processes. 
Such coupled on-chip resonators could be a promising basis for new types of on-chip optical switching, communication \cite{wangNovelLightSource2017a}, sensing \cite{zhongSensingExceptionalSurfaces2019} and machine learning application \cite{brunnerCompetitivePhotonicNeural2021, porteCompleteParallelAutonomous2021a, tirabassiBinaryImageClassification2022a}.

\subsection*{Materials and Methods} 

\subsubsection*{Coupled mode theory}
We model the coupled microdisk laser system around threshold through coupled mode theory. Coupled mode theory is an established theory for evanescently coupled microdisk lasers \cite{hodaeiParitytimeSymmetricMicroring2014, el-ganainyNonHermitianPhysicsPT2018, ozdemirParityTimeSymmetry2019, benzaouiaNonlinearExceptionalpointLasing2022a}.  We consider two single-mode resonators ($i = 1,2$) with resonant frequencies $\widetilde{\omega}_i = \omega_i - i \gamma_i$, where $\omega_i$ is the real frequency and $\gamma_i$ is the imaginary part corresponding to the loss. The resonators are coupled by a real coupling factor $\kappa$. We pump the resonators with pump $P_i$. When the system is around threshold we can neglect nonlinear gain saturation, and the steady-state behaviour is described by the linear 2 x 2 matrix eigenvalue equation,
\begin{equation}
    \left[
    \begin{matrix}
        \omega_1 - i \gamma_1 + i P_1
        & \kappa \\
        \kappa 
        & \omega_2 - i \gamma_2 + i P_2
    \end{matrix}
    \middle]
    \middle[
    \begin{matrix}
        \tpsi_1 \\ \tpsi_2
    \end{matrix}
    \middle]
    = \tomega
    \middle[
    \begin{matrix}
        \tpsi_1 \\ \tpsi_2
    \end{matrix}
    \right] \,.
    \label{CMTmain}
\end{equation}
The eigenvalues $\tomega$ of eq. \ref{CMTmain} are the modes of the coupled system, where $Re(\tomega)$ is the emission frequency and $Im(\tomega)$ is the the loss/gain of the system. For $Im(\tomega) > 0$ ($< 0$) the system is above (below) the lasing threshold of $Im(\tomega) = 0$.

For the above threshold solutions, Eq.~\ref{CMTmain} becomes nonlinear and we included gain saturation to find the solutions (see Supporting Information section SIII).

\subsubsection*{Microdisk Fabrication}

The bottom-up grown WZ InP microdisks are fabricated as reported in \cite{staudingerWurtziteInPMicrodisks2020a}.
The crystal growth and WZ structure are induced by 110-oriented trenches in a 300 nm \ce{SiO2} layer on a standard InP (111)A substrate. The trenches are fabricated using e-beam lithography (EBL) and reactive ion etching (RIE). InP is grown inside the trenches through metal-organic vapor phase epitaxy (MOVPE) resulting in fins, that are the base of the microdisks and define their diameter. Once the InP reaches the top of the trenches, zipper-induced epitaxial lateral overgrowth (ELO) extends the InP laterally over the \ce{SiO2} until the base of the hexagonal microdisk is formed. The growth then occurs vertically to form InP microdisks with flat crystalline sidewalls. 

\subsubsection*{Optical Measurements}


The InP microdisks are characterised in a micro photoluminescence setup. They are pumped with a 633 nm, fs-pulsed laser at 100 kHz. The pump beam is focused on the sample using a 100x objective. The microdisks are excited from above and their emitted light is also collected from above through the same objective. The emission spectra are measured in a Princeton Instruments spectrometer. 
We shape the pump beam into two half circles with independently adjusted powers by reflection off a DMD that consists of 1280 x 800 addressable mirror pixels. As single DMD pixels are not resolved due to their small size, turning off a fraction of the pixels reduces the power of the beam. To achieve uniform pixel distribution, we used Floyd-Steinberg's error diffusion dither algorithm incorporated in MATLAB. We measure the exact power on each disk with a calibrated power meter that measures light reflected off a glass slide. The glass slide is located before the objective and the ratio between the power at the powermeter and sample location was measured to calibrate the powermeter. To obtain the power on each disk separately the single beam halves were projected and the power obtained before the full pattern projection and spectrum measurement.

\subsection*{Acknowledgments}

AF, JD, HM, and KM acknowledge support from the EU ITN EID project CORAL (GA no. 859841).
TVR, DS, and RS acknowledge support from The Engineering and Physical Sciences Research Council (EPSRC), grant number EP/T027258.
WKN acknowledges the research support funded by the President’s PhD Scholarships from Imperial College London.
SV acknowledges support from EPSRC–EP/X52556X/1.
We thank Philipp Staudinger and the Cleanroom Operations Team of the Binnig and Rohrer Nanotechnology Center (BRNC) for sample fabrication together with HS.

\subsection*{Author Information}

\subsection*{Contributions}

Conceptualization: AF, RS, WKN, TVR, DS, JC; 
Optical Measurements: AF; 
Optical Setup: SV, DS, WKN, AF, JD; 
Theory: TVR, AF, JC, WKN; 
Data Analysis: AF, TVR; 
Fabrication: HS, KM; 
Visualization: AF, TVR, JC; 
Funding acquisition: RS, KM; 
Project administration: RS, KM; 
Supervision: RS; 
Writing (original draft): AF; 
Writing (review and editing): AF, TVR, RS, SV, DS, KM, HS, JD, WKN, JC.

\bibliography{ms}

\end{document}